\documentclass{sig-alternate-10pt}

\usepackage{cite}	% sort citation numbers
\usepackage{amsmath, mathptmx}
\usepackage{times}
\usepackage{subfig}
\usepackage{algorithm}
\usepackage{color}

\usepackage{url}

\usepackage{xspace}
\usepackage{graphicx}
\usepackage{paralist}

\usepackage{listings}
\long\def\com#1{}
\long\def\haoz#1{{\color{blue}{\bf Hao: }{\small [#1]}}}

\renewenvironment{itemize}{
   \begin{list}{\labelitemi}{
     \setlength{\topsep}{0ex}
     \setlength{\itemsep}{-0pt}
     \setlength{\itemindent}{0pt}
     \setlength{\leftmargin}{\labelwidth}
     \addtolength{\leftmargin}{-8pt}}
}{\end{list}}

\newcommand{\app}{NCVS\xspace}
\newcommand{\graph}{CDG\xspace}
\newcommand{\acq}{NSDMiner\xspace}

\newcommand{\ie}{{\em i.e.\xspace}}
\newcommand{\eg}{{\em e.g.\xspace}}

\newcommand{\para}[1]{{\smallskip\noindent {\bf #1}}}

\begin{document}

\title{\vspace{-1em}A Non-Intrusive and Context-Based Vulnerability Scoring 
Framework for Cloud Services}

\author{Hao Zhuang, Florian Pydde\\
\affaddr{EPFL}}

\maketitle

\begin{abstract}

Understanding the severity of vulnerabilities within
cloud services is particularly important 
for today's service administrators. 
Although many systems, \eg, CVSS, have been built to evaluate
and score the severity of vulnerabilities for administrators,
the scoring schemes employed by these systems fail to take into account
the contextual information of specific services having 
these vulnerabilities, such as what roles they play in a particular
service. Such a deficiency makes resulting scores unhelpful.
This paper presents a practical framework, \app,
that offers automatic and contextual scoring mechanism
to evaluate the severity of vulnerabilities 
for a particular service.
Specifically, for a given service $S$, 
\app first automatically collects $S$'s contextual information 
including topology, configurations, vulnerabilities and 
their dependencies.
Then, \app uses the collected information to build a 
contextual dependency graph, named \graph, to model $S$'s context.
Finally, \app scores and ranks all the vulnerabilities in
$S$ by analyzing $S$'s context, 
such as what roles the vulnerabilities play in $S$,
and how critical they affect the functionality of $S$.
\app is novel and useful, because 
1) context-based vulnerability scoring results are highly
relevant and meaningful for administrators to 
understand each vulnerability's importance specific to the
target service; and 2) the workflow of \app 
does not need instrumentation or modifications to any source code.
%In this paper, we present challenges, design, implementation 
%and preliminary evaluational results of \app.
Our experimental results demonstrate that \app can obtain
more relevant vulnerability scoring results than
comparable system, such as CVSS.

\end{abstract}

\section{Introduction}

Today's cloud-scale systems become increasingly complex -- they not only
employ multi-layered network/software stacks, 
but also deploy various distributed service components offered by
other providers. These structurally complex systems, 
nevertheless, may inadvertently introduce
much more vulnerabilities than traditional computer systems.%
\footnote{A vulnerability is a weakness (defect) of 
the software design or implementation rather than necessarily a bug;
in other words, a system may have a vulnerability due to
a defective design, even if its implementation is perfect.}
Therefore, it is important for administrators to 
score vulnerabilities based on their severity,
thus enabling administrators to deal with 
the critical ones accordingly.

Driven by the above motivation, many public vulnerability
databases (\eg, CVE~\cite{cve} and OSVDB~\cite{osvdb}) are maintained, 
which contain classification of vulnerabilities, 
description of the nature of their severities, as
well as the exploitability based on the feedback of security experts.
Furthermore, several vulnerability scoring systems are also
developed for the disclosure and severity ranking of vulnerabilities.
One of the most representative vulnerability scoring systems
is Common Vulnerability Scoring System (CVSS)~\cite{cvss}, 
which is designed based on expert knowledge, multi-dimension scoring
schemes and customer feedback. 

However, current vulnerability scoring efforts
typically quantify the severity of vulnerabilities 
in a general (or global) way -- \ie, 
how severe vulnerabilities are -- rather than 
considering the context of specific services 
holding these vulnerabilities. 
%In other words, they mainly consider the scale that they
%exist in all the services deployed equally.
Applying such scores to evaluate the risk of a particular service,
therefore, may not always be meaningful or instructive. 
For example, a denial of service vulnerability against MySQL 
might have a high severity score in CVSS. But for a particular
service context, its application components running MySQL
may only be used for exception event logging, which is rarely
invoked. Thus, the impact of this vulnerability on 
the whole service context would not be as significant
as the ranking score suggested in CVSS. 
%Thus, vulnerabilities specific to those systems might
%not be considered critical. 
On the contrary, if a service context has
to rely on such legacy systems for its core functionality,
their vulnerabilities should be ranked high and handled with
high priority.

Based on this insight, 
for a particular service administrator, it is
desirable to have a tool, which not only considers the
properties of vulnerabilities, but more importantly takes into
account the target service's context. 
Such a tool can offer a truly meaningful
severity score for each vulnerability to serve as a
guideline regarding which vulnerabilities are the most urgent
and critical ones for that {\em particular service},
thus enabling administrators to allocate
appropriate patches or fixing code to deal with them
accordingly.
Although many efforts (\eg, attack graph approaches%
~\cite{sawilla08identifying, huang11distilling, jiang12vrank, zhai15} 
and profiling techniques~\cite{zhao14iprof, zhaivldb17}),
these efforts are either {\em ad hoc} or hard to be
extended to tackle our problem (see $\S$\ref{subsec-related}
for more details).

This paper presents a systematic work, \app, 
which is a practical contextual vulnerability
scoring framework for cloud services. 
To evaluate the severity of vulnerabilities in terms
of a particular service context, \app first automatically 
collects contextual information about service components
and their dependencies in a comprehensive way. 
Using this acquired data, \app then builds
a {\em contextual dependency graph}, named \graph, to model
the service's context.
Finally, \app analyzes all the vulnerabilities' relative importance 
in this context and returns a contextual vulnerability scoring report.
To the best of our knowledge,
\app is the first effort capable of offering automatic (\ie,
non-intrusive) and context-aware vulnerability 
scoring for cloud services.

% the first challenge is how to comprehensively collect information

Building a non-intrusive system for contextual vulnerability scoring
needs to address several challenges.
First, the information of components in the target cloud service and
their associated dependencies should be comprehensively
acquired, since vulnerabilities may exist in different
types of components. In addition, infrastructures 
underlying today's services tend to be complex; thus,
asking an administrator to manually collect 
such a large dataset is an infeasible task.
Existing techniques {\em either} 
heavily rely on interruptive instrumentations (\eg, 
MagPie~\cite{barham04using} and 
Project~5~\cite{aguilera03performance}),
{\em or} have been developed for limited types of dependencies (\eg,
Sherlock~\cite{bahl07highly} and NSDMiner~\cite{natarajan12nsdminer,
peddycord12accurate}).
Thus, how to {\em comprehensively} acquire the detailed 
contextual dependencies (including 
network, hardware and software-level dependencies) 
{\em without instrumentations} presents a challenge.
We develop a non-intrusive approach capable of
collecting three types of contextual information by 
mining log information and network traffic (detailed in
$\S$\ref{subsec-collection}).

% the second challenge is how to model and reason about vul
The second challenge includes 1) how to model the context with 
the collected information and 2) how to compute the importance of 
each vulnerability by taking into account this context.
We propose a new representation, 
named contextual dependency graph (or \graph),
to model the target service's context.
In general, \graph is a two-layered 
directed acyclic graph (DAG) representation,
and is capable of capturing more details than
previous dependency graph models. With \graph in hand, 
we propose a pluggable scoring module that allows 
administrators to apply diverse (existing or customized) 
ranking algorithms based on their requirements.
As an example, we also propose a new algorithm specific to
our \graph model by adapting PageRank algorithm~\cite{page1999pagerank}.
Our experimental results indicate the new algorithm can
obtain more reasonable vulnerabilities' scores than CVSS (detailed in
$\S$\ref{subsec-scoring}).  

In summary, this paper mainly makes four contributions.
First, we present the first practical (\ie, general, non-intrusive,
pluggable and effective) contextual vulnerability scoring framework
for cloud services. 
Second, we develop a set of collectors capable of automatically
acquiring a target cloud service's contextual information,
including network, service and hardware dependencies.
% an automatic contextual
%information collection system that can acquire comprehensive dependency
%information underlying services of interest.
Third, we construct a new graph model, named \graph, specific to
our collected information and vulnerability scoring purpose.
Finally, we demonstrate the practicality of \app based on
a lab-scale case study and performance measurements.

\section{Motivation and Related Work}
\label{sec-motiv}

This section first motivates our work
($\S$\ref{subsec-motivating}),
and then discusses 
why existing work does not help ($\S$\ref{subsec-related}).

\subsection{Motivating Example}
\label{subsec-motivating}

Figure~\ref{fig-motivation} presents a simple but illustrative example
for our motivation. 
In this example, we consider a real-world
lab-scale Hadoop cluster with eight
servers: one name node (S1), one backup name node (S5),
and six data nodes (S2-S4, and S6-S8).
The name nodes and data nodes run job trackers and
task trackers, respectively. 
S1-S4 and S5-S8 belong to two individual racks, respectively.
Each rack has one edge switch~\cite{zhaip2p09, zhaippna}.

By retrieving one of the most representative vulnerability scoring
systems, Common Vulnerability Scoring System (CVSS)~\cite{cvss},
the administrator discovers 
two vulnerabilities in this cluster:
1) Hadoop name node vulnerability, CVE-2015-7430, with severity score 8.4; 
and 2) core switch vulnerability, CVE-2016-1392, with severity score 7.4.  
In the example of Figure~\ref{fig-motivation}, the former exists in
Hadoop components running in S1 and S5, while the later exists in
Core1 switch software. 
If this administrator prioritizes her fixing or patching vulnerabilities 
based upon the scores provided by CVSS, she should fix Hadoop
vulnerability first. 
Nevertheless, if we deeply analyze the given Hadoop cluster, 
we may find that the states of name nodes in this Hadoop cluster have been 
replicated across two servers (\ie, S1 and S5).
Even though the vulnerability CVE-2015-7430 is triggered on one name
node (\eg, S1), the service can use S5 -- the backup name node -- to 
manage all data nodes. 
On the contrary, if the vulnerability CVE-2016-1392
is triggered, any network traffic from external network cannot reach any
servers in this cluster.

{
\begin{figure}[tbp] \centering
\includegraphics[width=0.45\textwidth]{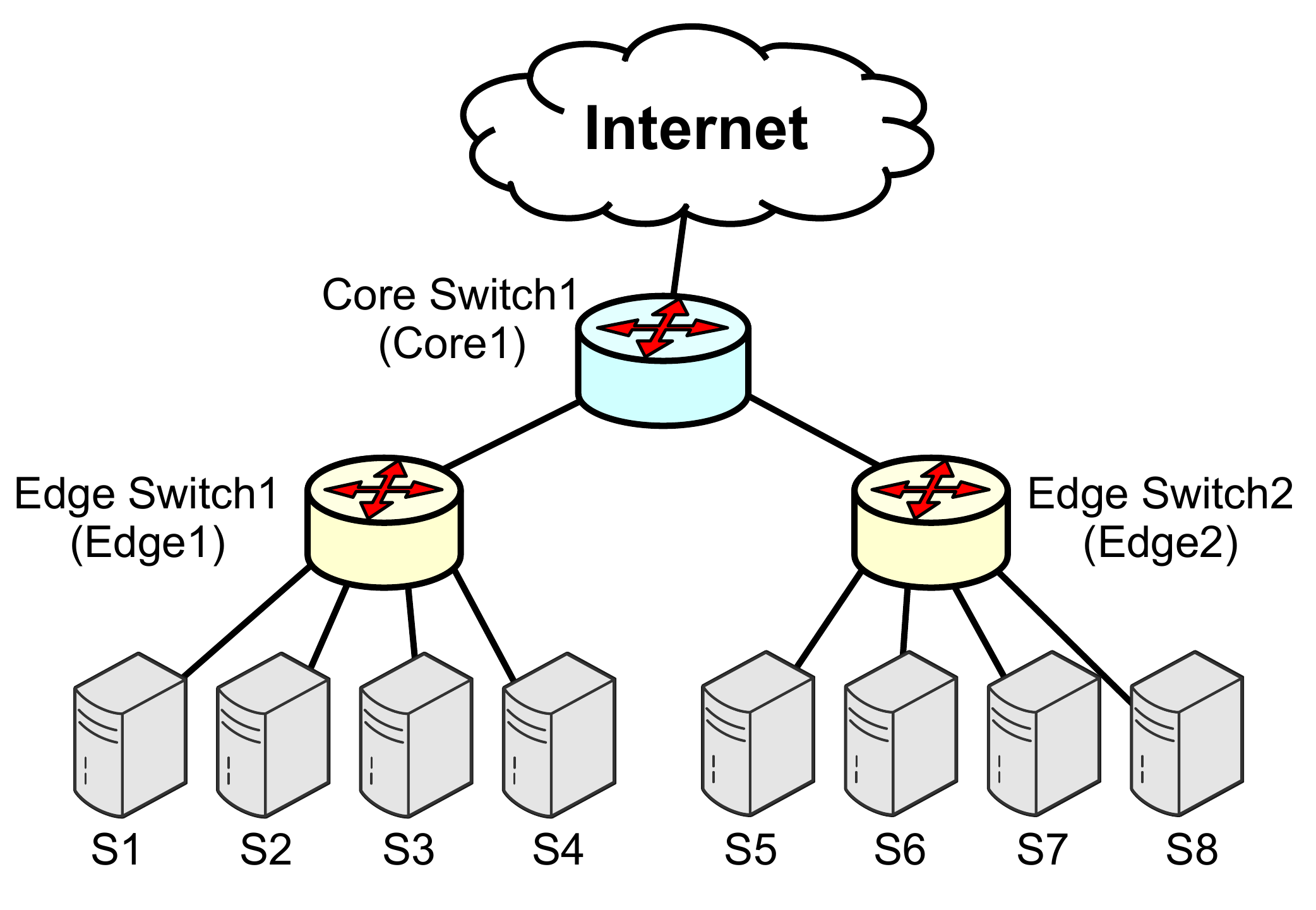}
\caption{Motivating example.}
\label{fig-motivation}
\end{figure}
}

Based on the above example, we note that existing vulnerability
scoring systems fail to guide administrators to understand
the severity of vulnerabilities in their service deployments.
In other words, because existing vulnerability scoring systems do not 
take into account contextual information of target vulnerabilities,
severity scores offered by these systems are not truly helpful in
practice. Thus, we raise a question that whether it is possible to
build a practical contextual vulnerability scoring system?

This paper develops such a system, named \app.
In Figure~\ref{fig-motivation} example, if the administrator uses \app
to score the two revealed vulnerabilities, she can obtain a different
result: CVE-2016-1392 has a higher severity score than 
CVE-2015-7430 in terms of service context as shown in
Figure~\ref{fig-motivation}.

\subsection{Existing Efforts Discussions}
\label{subsec-related}

We now describe and compare existing efforts,
and discuss why they do not work for our purpose.

\para{Vulnerability scoring systems.}
Many current vulnerability scoring systems (\eg, CVSS~\cite{cvss}, 
OSVDB~\cite{osvdb} and OVAL~\cite{oval}) have been developed to score
vulnerabilities based on their severity.
However, these systems only score vulnerabilities generally, \ie, 
how severe they are by themselves
without considering any specific environment.
Such a score 
%is indeed reasonable for security experts when they
%aim to understand the impacts of vulnerabilities, 
%but it 
is not meaningful or instructive to evaluate vulnerabilities
in specific service context (like the example in
$\S$\ref{subsec-motivating}). 

Although several contextual vulnerability scoring efforts,
\eg, improved CVSS~\cite{fruhwirth09improving, cvss3} and  
VRank~\cite{jiang12vrank}, have been proposed,
they are quite {\em ad hoc} and impractical.
First, these efforts do not have any automatic environmental information 
collection capability. This means all the contextual information 
has to be input manually, making their scoring process 
impractical to any complex services. 
Second, these approaches only support coarse-grained scoring
schemes, because they lack expressive models. 
Finally, the environmental factors considered by these efforts
are too simple to represent the contexts of
real-world services.

\para{Attack graph based risk evaluation efforts.}
Attack graph based risk reasoning 
approaches~\cite{ou05mulval, ou06ascalable, huang11distilling,
sawilla08identifying} have been studied in the
past ten years. An attack graph is an abstraction of 
the details of possible attacks against any computer system or service.
Attack graph based efforts are appropriate to evaluate the 
potential risks of target services, rather than computing 
vulnerabilities' severity.
In other words, attack graph techniques leverage existing vulnerability 
scores (\eg, provided by CVSS) to compute the risks of services of 
interest, but it is not capable of updating vulnerabilities' scores
based on different services' contexts.

\section{\app Design}

This section first presents a high-level 
overview of \app framework in $\S$\ref{subsec-overview}.
Then, we detail two important modules in $\S$\ref{subsec-collection}
and $\S$\ref{subsec-scoring}, respectively.

\subsection{\app Framework Overview}
\label{subsec-overview}

As shown in Figure~\ref{fig-overview},
\app has two important modules: 
1) contextual information acquisition module and 
2) vulnerability scoring module.
In addition, \app relies on two databases -- contextual
information DB and vulnerability threats DB -- that are used
to provide necessary information for vulnerability scoring.

%As depicted in Figure~\ref{fig-overview},
\app performs the following two steps to score vulnerabilities for
a target cloud service $S$.

\para{Step~1:}
\app's contextual information acquisition module automatically collects 
comprehensive dependency information, including topology, service
deployment, and correlations between components.
Then, the module stores the collected information
into the contextual information DB for post processing.
All the operations within this step do not require any 
additional instrumentations from administrators.
$\S$\ref{subsec-collection} discusses this module's
design in more detail.

{
\begin{figure}[tbp] \centering
\includegraphics[width=0.45\textwidth]{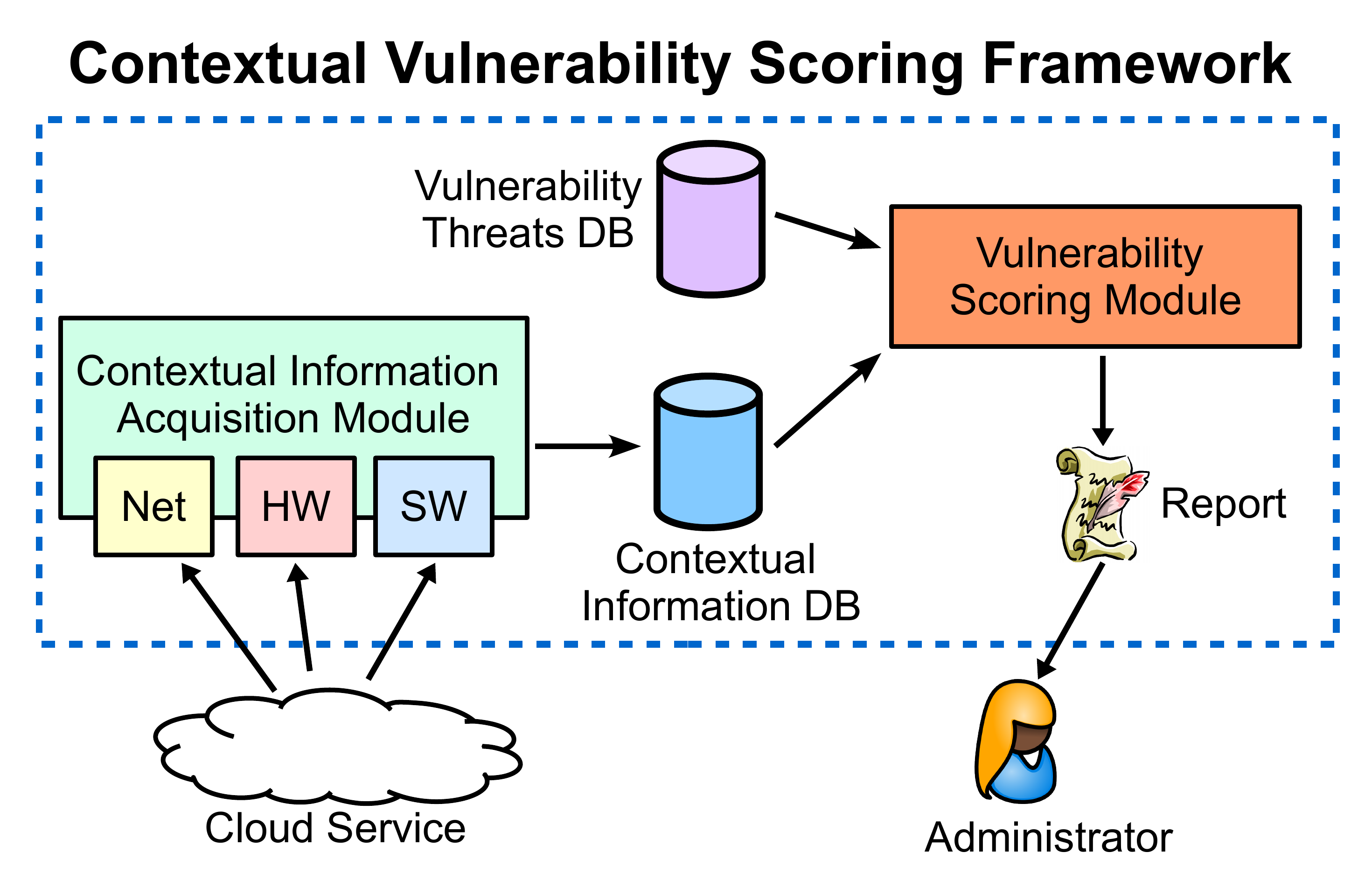}
\caption{\app's architecture overview.}
\label{fig-overview}
\end{figure}
}

\para{Step~2:}
The vulnerability scoring module takes the information in 
the contextual information DB and vulnerability threats DB as input, 
and constructs a dependency graph, named \graph, 
modeling the context of the target service $S$.
While various graphs representing service structures have been proposed
in the past years,
\graph is the first effort capable of capturing comprehensive
contextual information.
With the \graph in hand, the module analyzes it and 
outputs a report ranking vulnerabilities based on their
scores specific to the context of the target service $S$.
$\S$\ref{subsec-scoring} details this module.

% This is the sub-section describing collection module
\subsection{Contextual Information Acquisition}
\label{subsec-collection}

For a target service $S$,
\app's contextual information acquisition module automatically collects
three types of dependency information (network dependency, 
hardware dependency and software dependency),
and then stores the collected information into 
the contextual information database (as shown in
Figure~\ref{fig-overview}).
We focus on the collection of 
these three types of contextual information, 
since this information is able to comprehensively 
represent a service's context (or environment)~\cite{zhai14heading,
zhai13}.
To our knowledge, there is no existing effort
that can simultaneously collect these three types of dependency
information without any instrumentation.

%\begin{description}

{\em Hardware dependency} means $S$'s physical 
  topology information, 
  including the locations of servers and network
  devices (\eg, aggregation switches and core routers) as well as
  the links between servers and network devices.
  In other words, hardware dependency information is the data-center 
  infrastructure components and their links used to support service
$S$~\cite{zhaiapweb11}.

{\em Software dependency} denotes software components running on
  each node (\eg, server or switch). For switches, the software
  components might be the switch OS and related 
  embed functional applications; for servers, the software components
  include operating systems, service $S$'s components (\eg, 
  back-end DB), and needed applications (\eg, SSH)~\cite{wang12,
  sun11, zhai11, wangfskd10, wangicdpads10, liu09, zhaisocialcom09}. 

{\em Network dependency} means an invoking sequence (or flow) between 
  different software components in the target service $S$.
  For example in a MapReduce service, a request is first sent 
  to a job tracker and then arrives at several task trackers,
  forming a directed invoking flow. Such an invoking sequence,
  \eg, Client -> JobTracker -> TaskTrackers, represents an item
  of network dependency information.
  Different from hardware dependency, network dependency is at logical
  level, which means a network dependency (\ie, an invoking flow) may 
  be across different servers or only occurs on the same
  server. 

As described in the introduction section,
the first challenge of building \app is how to
acquire all of the above three types of information 
without any instruments or modifications to the source code
of target services.
We now describe how we develop 
the contextual information acquisition module.

%\subsubsection{Automating Contextual Information Collection}

%As described in the introduction section,
%the first challenge of building \app is how to automatically 
%acquire the three types of contextual information.

\para{Software dependency collection.}
The basic idea of software dependency collector is to determine software
components running on each node (\eg, server or switch) 
by analyzing the node's log information during
a service request session. A software dependency is defined as 
{\tt <node="M" sw="S" dep="x,y,z"/>}, where {\tt node} indicates 
the node running (or holding) this software component specified 
by {\tt sw}, and {\tt dep} shows 
all the components (\eg, libraries) used by
this software component {\tt sw}.

Our insight for designing the software dependency collector
is: when a request is sent to the target service $S$, 
each involved node would generate log corresponding 
to its functions, and the log information
across all the nodes has a time-aware sequence.
Thus, if we can capture and analyze all the involved nodes'
log information between the start time and end time of accessing $S$,
we would be able to extract the software dependency information
of each node supporting $S$. In our design, therefore,
the software dependency collector -- playing as a client role -- sends 
many requests to the target service $S$ 
and infers software dependencies of each
node by analyzing the generated log during the session.
The collector employs a well-known association rule mining algorithm, 
named Apriori~\cite{agrawal93mining}, 
that first groups each item of log information by a time interval,
then mines correlated rules among log items with confidence values,
and finally selects these correlations with 
the high confidence level as software dependencies. 
%then extracts feature information by filtering noise,
%and finally determines software dependencies.
%to analyze the relationships among
%the generated log information and organize them as a dependency
%graph structure.
The accuracy of collector is related to the threshold of time interval
we set -- smaller threshold gives better accuracy.

\para{Network dependency collection.}
In order to automatically acquire needed network dependency information,
we design a network dependency collector based on the similar 
intuition as the software dependency collector.
In particular, we define a network
dependency as a continuous, directed stream of packets 
between two services (or applications).
Similar to the software dependency collector,
the network dependency collector also sends many requests to the 
target service $S$, but it aims to capture all the networking
packets and flows during the service access session.
To capture these traffics, our collector employs 
\acq~\cite{peddycord12accurate, natarajan12nsdminer},
which is a well-developed traffic monitor.
%than Sherlock~\cite{bahl07highly} and Orion~\cite{chen08automating}.
An important reason we choose \acq is that 
\acq does not need to install extra agents or software on involved
nodes; moreover, \acq can collect more accurate traffic,
thus making our network dependency collector work better than
Sherlock~\cite{bahl07highly} and Orion~\cite{chen08automating},
two representative network dependency collection tools.
%2) \acq is an open source tool written in Python~\cite{nsdminer}
%and its output format is well suitable to \app.
The network dependency collector 
outputs many items of the network dependency information
formatted as follows:
\lstset{
  language={[Sharp]C},
  basicstyle=\ttfamily\footnotesize
}
\begin{lstlisting}[language=C]
<IP address A>:<port A>:[TCP] <# App~1>
  <IP address B>  <port B>  [TCP]
  <IP address D>  <port D>  [TCP]
<IP address B>:<port B>:[UDP] <# App~2>
  <IP address C>  <port C>  [UDP]
  <IP address D>  <port D>  [UDP]
...
\end{lstlisting}

This example shows that component $A$ which runs on
IP address $A$ at port $A$ depends on components $B$
and $D$ which run on their respective IP addresses
and ports. Similarly, component $B$ depends on component $C$ and $D$.
Note that all the components ($A, B, C$ and $D$) are 
services or applications.

\para{Hardware dependency collection.}
Since most of current data-center networks employ software defined 
network (SDN) controller~\cite{mckeown08openflow} or network provenance
techniques~\cite{zhou11secure, zhou11nettrails}, 
it is easy for the administrators to learn
physical network topology (\ie, hardware dependency information
for our system). Thus, if an administrator has known network
topology, our hardware dependency collector would directly 
read and parse the network topology file that might be
written in different formats (\eg, CSV or XML), thus obtaining
our needed hardware dependency information.

In the case that there is no such file available,
we also provide a network topology generation toolkit
(TopoGen) 
to assist administrators to derive the network topology in a fast, 
less tedious way. TopoGen supports various data-center topology
models including fat tree~\cite{mysore09portland} 
and BCube~\cite{guo09bcube}. The administrators only
need to give the basic information, \ie, the number of
core/aggregation/edge switches, to TopoGen,
and then TopoGen generates a network topology
automatically. Administrators are allowed to validate the auto-generated
topology and further modify if necessary. Furthermore, TopoGen allows
users to plug new topology generation rules into generate network links,
which is very efficient and helpful for large-scale network systems.
Thus, combining with our TopoGen toolkit and the basic information from
cloud administrators, we can derive accurate network topology
information in a less tedious manner.

% This is the sub-section describing vulnerability scoring module
\subsection{Vulnerability Scoring}
\label{subsec-scoring}

Vulnerability scoring module computes the contextual
score of each vulnerability in $S$ by two steps:
1) modeling $S$'s context with the contextual information
acquired in $\S$\ref{subsec-collection}, and 2) computing
the importance of each vulnerability in terms of this model.

\subsubsection{Contextual Dependency Graph (\graph)}

The vulnerability scoring module first needs to build an explicit
graph, named contextual dependency graph (or \graph),
to represent the context of the target service $S$.
Figure~\ref{fig-cdg} presents a \graph example.

In a \graph, there are two layers: hardware (physical) layer and software
(logical) layer. All the nodes representing physical components (\eg,
servers and switches) are involved in the hardware layer;
on the contrary, all the nodes representing software components (\eg,
libraries and applications) are put in the software layer.
All the edges in the \graph are directed. An edge like $A \rightarrow B$
means component $A$ depends on another component $B$.

Because our contextual information acquisition module 
automatically collects hardware, software and network dependency
information, we now describe the correlation 
between the collected contextual information and a \graph.
All the hardware components (\eg, servers and switches)
and their dependencies (\eg, links)
are modeled as \graph nodes in the hardware layer and directed
edges among them, respectively. 
In other words, a directed edge between two \graph nodes
in the hardware layer means a physical topology link.
All the software dependencies of each machine are modeled as 
\graph nodes in the software layer, and they are connected by directed
edges to their corresponding nodes in the hardware layer,
as shown in Figure~\ref{fig-cdg}.
Namely, a directed edge across the hardware layer and the
software layer in a \graph means a software dependency.
All the network dependencies are modeled as directed edges 
connecting different \graph nodes in the software layer.
Namely, a directed edge between two \graph nodes in the software layer
means a network dependency.

{
\setlength{\abovecaptionskip}{-.2pt}
\setlength{\belowcaptionskip}{-10pt}
\begin{figure}[tbp] \centering
\includegraphics[width=0.44\textwidth]{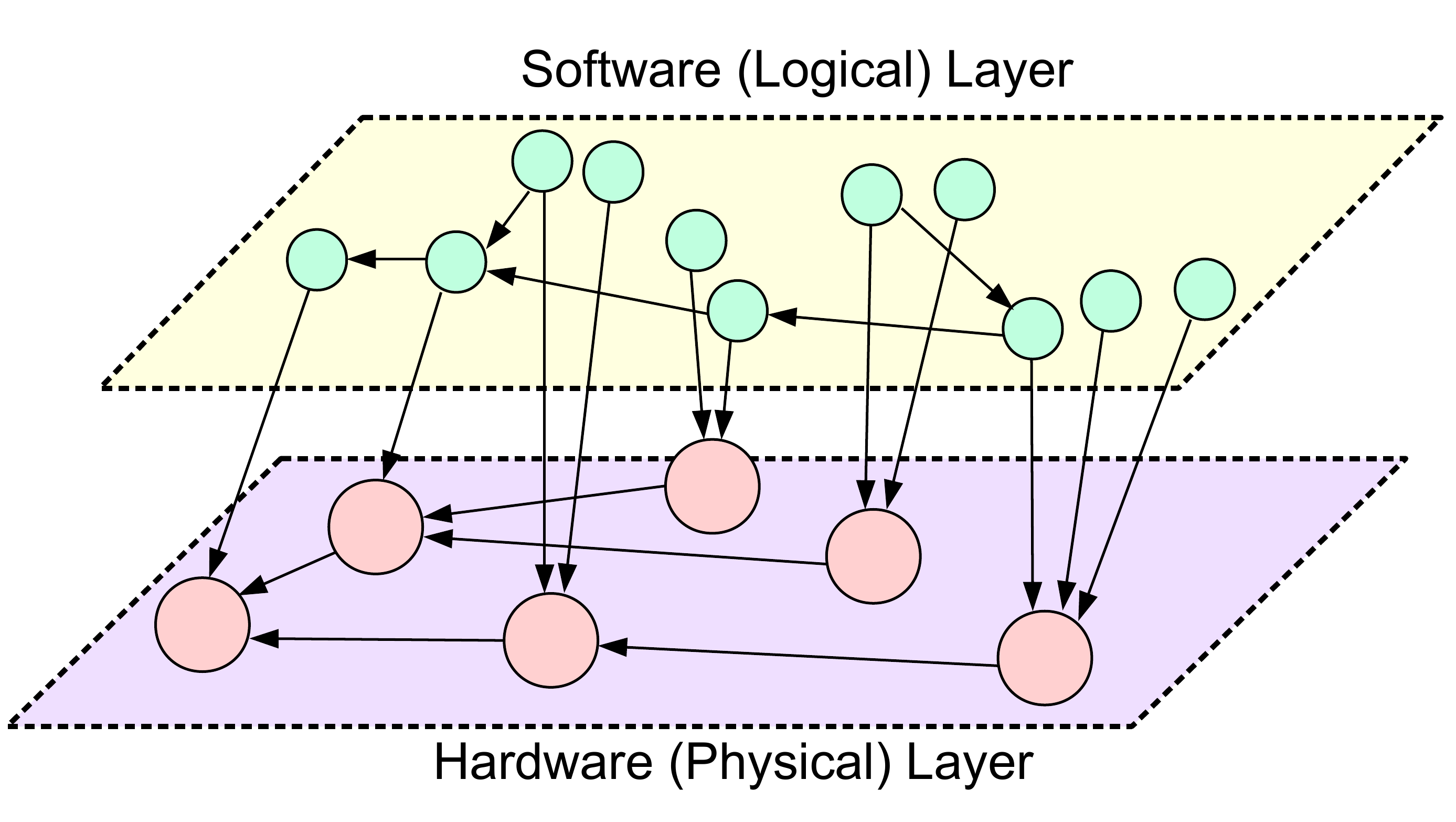}
\caption{Contextual dependency graph (\graph).}
\label{fig-cdg}
\end{figure}
}

\subsubsection{Vulnerabilities' Scores Computation}

With a \graph in hand, the vulnerability scoring module now aims 
to identify what nodes have vulnerabilities in the \graph and
score these identified vulnerabilities by taking into account
their ``roles'' in this \graph.
As shown in Figure~\ref{fig-overview},
the vulnerability scoring module relies on another database,
named vulnerability threats DB.
The vulnerability threats DB could be any existing
vulnerability scoring system -- our prototype employs CVSS.
We need such a database because it can offer exploited vulnerabilities
and their basic threats given by security experts.

\para{Identify involved vulnerabilities.}
%There are many open databases to access CVE vulnerability data. 
The vulnerability scoring module first uses the node information in \graph
to search vulnerabilities involved in the service $S$'s deployment.
In our prototype, we implemented keyword matching and classification
methods to identify involved vulnerabilities in CVSS, 
since CVSS provides detailed information about each recorded
vulnerability such as its software name, threat, and impact.
Note that because almost of all the vulnerability databases store
software vulnerabilities, \app also focuses on software vulnerabilities
rather than hardware vulnerabilities.

\para{Node ranking by considering context.}
After determining the involved vulnerabilities in the \graph, 
we start to score nodes and vulnerabilities in our target context. 
Specifically, for each affected software node $n$ in \graph, 
we compute its importance in terms of three different 
sub-graphs in \graph:

\begin{itemize}
 
\item Topology-aware importance $ti(n) = Rank(hw\_graph, n)$, where 
  $hw\_graph$ means the nodes that are located at the hardware layer 
  and have software component $n$.
\item Software-aware importance $si(n) = Rank(sw\_graph, n)$, where 
  $sw\_graph$ means the nodes that are located at the software layer 
  and have software component $n$.
\item Network-aware important $ni(n) = Rank(net\_graph, n)$, where 
  $net\_graph$ means the nodes that are involved in network dependencies 
  and have software component $n$.

\end{itemize}

In the above computation, $Rank()$ means graph node ranking function.
In our prototype, we adopt PageRank algorithm~\cite{page1999pagerank} 
to enable the graph node ranking function, \ie,
computing the importance of the node $n$.
Our prototype supports any graph node ranking algorithm,
including PageRank, HITs~\cite{kleinberg98authoritative} 
and new developed ranking algorithms.

\para{Ranking vulnerabilities.}
With the importance of each software node in hand,
we can compute the contextual scores for vulnerabilities involved in
our target service $S$. 
In particular, for a given vulnerability $v$,
we compute the score of $v$, $severity(v)$, as follows.
\[
severity(v) = w_{ti} * \sum_{i=1}^{|S_v|}ti(n_i) 
  + w_{ni} * \sum_{i=1}^{|S_v|}ni(n_i) 
  + w_{si} * \sum_{i=1}^{|S_v|}si(n_i)
\]

Where, $S_v$ means the set of software nodes that contain the
vulnerability $v$. $w_{ti}$, $w_{ni}$ and $w_{si}$ are used to
weight the three types of context impacts, respectively.
These weights are first assigned based on expert knowledge and further
tuned in an iterative process. For example, we can start with equal
importance for three contexts $w_{ti}=w_{ni}=w{li}=1$ and adjust these
weights to control the score until the final scores are reasonable.  
The design of the aggregation function is also pluggable, 
which means developers can customize their aggregate function 
according to their particular purpose.  
For example, in our prototype, we also provide another aggregate function 
to integrate CVSS score based on the product rule: $severity(v) =
(\sum_{i=1}^{|S_v|}ti(n_i)*ni(n_i)*si(n_i)) * CVSS(v)$. 
In this aggregate function, CVSS
score is considered as a local software importance factor.

\section{Evaluation}
\label{sec-eval}

We have developed a prototype system in Java to
evaluate \app. In this section,
we first evaluate \app's performance 
($\S$\ref{subsec-bench}), and then evaluate the effectiveness
of \app with a real-world case study ($\S$\ref{subsec-case}).

\subsection{Performance Evaluation}
\label{subsec-bench}

We deploy our \app prototype on a workstation equipped with 
Intel Xeon E5-2630 2.30 GHz CPU and 64 GB RAM.
For performance evaluation, we measure the performance of 
two modules of our \app prototype -- \ie, the runtime for
two modules to handle different sizes of services.

First, we evaluate the runtime of the first module of \app,
\ie, contextual information acquisition module.
The performance bottlenecks of this module are
software dependency collector and network dependency collector,
so that our experiments focus on these two collectors.

{
\setlength{\belowcaptionskip}{-7pt}
\begin{figure}[t]
\center
\includegraphics[scale=0.44]{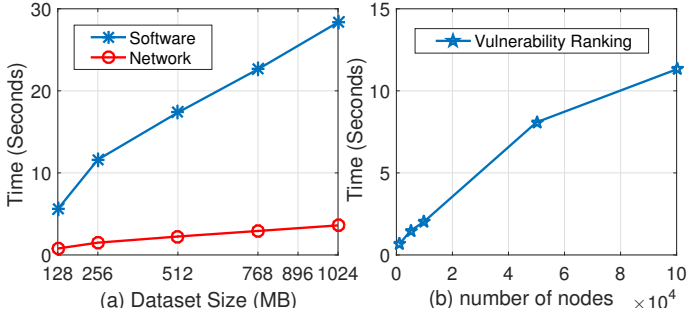}
\caption{Performance evaluation: 
(a) contextual information acquisition module,
and (b) vulnerability scoring module.}
\label{fig:timecost}
\end{figure}
}

Figure~\ref{fig:timecost}(a) presents the evaluational results 
for software and network dependency collectors.
In particular, we vary the size of target services, thus
getting raw datasets with different sizes (reflected by
the x-axis of Figure~\ref{fig:timecost}(a)). 
Then, we measured how long two modules 
need to extract dependency information of interest.
As shown in Figure~\ref{fig:timecost}(a),
the runtime of two collectors increases with the dataset size, 
but the software dependency collector is
more expensive due to the complexity of Apriori algorithm. 
It is worth mentioning that the increase in runtime over the size of 
services is acceptable in practice, since such a process is typically 
operated off-line and the results could be re-used in the future.

On the right hand, Figure~\ref{fig:timecost}(b) shows that the runtime
of vulnerability ranking algorithm (the performance bottleneck
of the vulnerability scoring module) increases with the number of nodes in
the CDG. We can observe that \app is quite efficient, with less than 12
seconds ranking 100,000 nodes.  Our ranking algorithm is powered by
PageRank algorithm which is highly efficient for large-scale graphs
(here we set the maximum number of iteration is 100 and the error
tolerance of two consecutive iterations is 0.001).

\subsection{Case Study}
\label{subsec-case}

In order to evaluate the effectiveness of \app,
we construct a real-world case study similar to our motivating
example (in $\S$\ref{subsec-motivating}).
In the case study, we deploy a Hadoop service on
a lab-scale cluster with 16 server machines and 8 switches. 
The \graph generated by \app to model this service contains 
747 nodes and 1204 dependencies. 
As shown in Table~\ref{tab:cvss},
we extract ten vulnerabilities from this service 
and derive their scores from CVSS (the fourth column in
Table~\ref{tab:cvss}). We observe 
that the vulnerabilities 2, 4, 5, 9 are network-related and located in
the switches, while the remaining ones are vulnerabilities of Hadoop 
and its related libraries (\eg, Apache common library). 
The fifth column in Table~\ref{tab:cvss} 
shows these ten vulnerabilities' scores produced by \app.
%In the severity computation, 
%we set the weights (\ie, $w_{ti}$, $w_{ni}$ and $w_{si}$) as 1,
%and use PageRank as the node ranking algorithm.

\begin{table}[t]
\centering
\caption{Comparing the vulnerability scores output
by CVSS and \app in our case study. The fourth column
in this table contains scores produced by CVSS,
while the fifth column includes \app's scores for our 
evaluated vulnerabilities.}
\begin{small}
\begin{tabular}{|l|c|c|c|l|}
\hline
Seq & ID            & Type           & CVSS & \app   \\ \hline\hline
1   & CVE-2015-1776 & Hadoop        & 6.3  & 0.0257 \\ \hline
2   & CVE-2015-4279 & Network            & 7.8  & 0.1069 \\ \hline
3   & CVE-2015-5210 & Hadoop         & 5.8  & 0.0275 \\ \hline
4   & CVE-2015-6355 & Network           & 5.0  & 0.0403 \\ \hline
5   & CVE-2015-6415 & Network            & 7.1  & 0.0770 \\ \hline
6   & CVE-2015-6420 & Hadoop  & 7.5  & 0.0287 \\ \hline
7   & CVE-2015-7430 & Hadoop         & 8.4  & 0.0293 \\ \hline
8   & CVE-2016-0731 & Hadoop         & 4.9  & 0.0269 \\ \hline
9   & CVE-2016-1503 & Network          & 9.8  & 0.0789 \\ \hline
10  & CVE-2016-2170 & Hadoop & 9.8  & 0.0303 \\ \hline
\end{tabular}
\end{small}
\label{tab:cvss}
\end{table}

In Table~\ref{tab:cvss}, we observe that 
if we fix these vulnerabilities based on the scores output by CVSS, 
the order should be 10 -> 9 -> 7 -> 2 -> 6 -> 5 -> 1 -> 3 -> 4 -> 8. 
On the contrary, if we fix vulnerabilities according to the
scores output by \app, 
the order should be 2 -> 9 -> 5 -> 4 -> 10 -> 7 -> 6 -> 3 -> 8 -> 1. 
To demonstrate \app's effectiveness, we construct
an interesting ``vulnerability fixing'' experiment. 
We assume that the current cluster is compromised by these ten 
vulnerabilities and all 16 server machines are not alive. 
Now, we fix these vulnerabilities based on the 
above two orders output by CVSS and \app, respectively. 
At each time point $t$, we only fix one vulnerability. 
After each time point, we evaluate the system in terms of the number of 
alive nodes in the cluster (as a safety metric). 
Figure~\ref{fig:comparision} shows our experimental results.  
We observe that fixing vulnerabilities according to the scores output
by \app can recover the capability of the Hadoop cluster
much faster than fixing vulnerabilities based on CVSS ranking list.
Thus, we can say \app can output more meaningful vulnerabilities'
scores than CVSS since \app takes into account the context
of the target service.
%the Hadoop cluster much faster 
%than that only based on CVSS ranking list. 
%The simulated results indicate that our \app is effectively to give 
%high rankings to these vulnerabilities that affect important nodes
%in the target service. 

{
\begin{figure}[t]
\center
\includegraphics[scale=0.48]{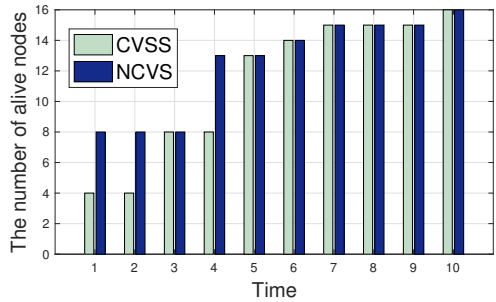}
\caption{Comparison of two different fix plans.}
\label{fig:comparision}
\end{figure}
}

%Compared with the order list that is ranked based on CVSS score, 
In addition, \app's results have the following advantages.
First, the result is network topology aware. 
We observe that network-related vulnerabilities in \app's ranking 
are generally higher than that of CVSS ranking list.
Second, the result can distinguish the roles of different software
components. We observe that the vulnerability (\ie, CVE-2015-7430) 
that affect master nodes ranks higher than 
that that affect slave nodes (\ie, CVE-2015-1776).
Finally, the result is accurate. In the case that two vulnerabilities 
are located in the same software component (\eg,
CVE-2015-6420 and CVE-2016-2170), their severity ordering
depend on their scores in CVSS, which means \app does not violate
the basic severity of each vulnerability.  
%Thus, we conclude that our ranking with \app 
%is effectively to rank vulnerabilities with considering their contexts. 

\com{
\subsection{Discussion}
One practical issue of our system is that determining appropriate weights for different types of contexts in \app is an iterative improvement process. This suggests that cloud administrators have to assign the initial values of weights based on their experience and finally find the suitable weights by trial and error.  In the real environment, this process could take a long time, which may also incur high costs in monetary and performance (\ie service availability). The main reason for this is that we lack of ground truth to evaluate the goodness of ranking lists. 
However, our analysis based on simulation in \ref{sec:casestudy} provides an easy and economical way for cloud administrators to speed up this process.  Our current work also focuses on providing insights for determining weights with a large-scale experimental evaluation. 

}

\section{Conclusion}

In this paper, we propose, design, implement and evaluate a novel
framework, \app, that can automatically score vulnerabilities 
for cloud services. Different from existing efforts, \app's
workflow is non-intrusive and \app scores
any given vulnerability by not only considering its intrinsic threats,
but also taking into account its service context.
Our evaluational results demonstrate \app's effectiveness (by
comparing with CVSS) and performance.

%For the next step, we intend to improve \app by proposing more
%sophisticated ranking algorithm and constructing better model
%to capture more comprehensive information.

%\begin{small}
\bibliography{os,net}
\bibliographystyle{plain}
%\end{small}

%\appendix
%\input{append}

\end{document}